# hybrid-drive pressure suppressing implosion instabilities and offering nonstagnation hotspot ignition with low convergence ratio for high-gain inertial fusion

Jiwei Li[1,2], X. T. He[1,2], Lifeng Wang[1,2], Yaohua Chen[1], Yan Xu[1], Bin Li[1], Minqing He[1], Hongbo Cai[1,2], Liang Hao[1], Zhanjun Liu[1,2], Chunyang Zheng[1,2], Zhensheng Dai[1], Zhengfeng Fan[1,2], B. Qiao[2,3], Ji Yan[4], Fuquan Li[4], Shaoen Jiang[4] and Shaoping Zhu[1,2]

[1] Institute of Applied Physics and Computational Mathematics, P. O. Box 8009, Beijing 100094, China

[2] Center for Applied Physics and Technology, HEDPS, and College of Engineering, Peking University, Beijing 100871, China

[4] School of Physics, Peking University, Beijing, 100871, China

[3] Center for Laser Fusion, China Academy of Engineering Physics, Mianyang 621900, China

E-mail: xthe@iapcm.ac.cn 21 August 2020

In laser-drive ICF, hybrid drive (HD) combined direct drive (DD) and indirect drive (ID) offers a smoothed HD pressure $P_{HD}$, far higher than the ablation pressure in ID and DD, to suppress hydrodynamic instabilities. In this letter, simulations of a new robust HD ignition target show that maximal HD pressure as high as $P_{HD} \sim 650$ Mbar driven by a novel "bulldozer" effect is achieved, resulting in nonstagnation hotspot ignition at the convergence ratio $C_r \approx 23$, and finally fusion energy gain~10 in total laser energy $= 1.42$ MJ. Two-dimensional simulations have confirmed that hydrodynamic instabilities are suppressed. A well-fitted scale of maximal HD pressure $P_{HD}(\text{Mbar}) = BE_{DD}^{1/4}T_r$ is found from simulations of different targets and laser energies as long as $T_r > 160$ eV , where B is the constant depending on ablator materials, $E_{DD}$ in kJ is DD laser energy and $T_r$ in 100eV is radiation temperature depending on ID laser energy $E_{ID}$. $P_{HD} \geq 450$ Mbar is requested for hotspot ignition. This scale from "bulldozer" effect is also available as $E_{DD}$ is reduced to ~kJ. Experiments have verified $P_{HD}$ about 3.5 times radiation ablation pressure for CH ablator using $E_{ID} = 43$ kJ ( $T_r \approx 200$ eV) and $E_{DD} = 3.6$ kJ, also shown that both backscattering fraction and hot electron energy fraction for DD laser intensity $\sim 1.8 \times 10^{15}$ w cm$^{-2}$ are about a third of the traditional DD laser-plasma interaction.

In the laser-drive inertial confinement fusion (ICF), the indirect drive (ID) approach [1] and direct-drive (DD)[2] approaches have widely been applied to investigate ICF ignition. In the ID approach, the shaped lasers act on an inner wall of a high-Z-shell hohlraum and its energy is converted to thermal x rays with radiation temperature T$_r$ . Thermal x rays ablate the outer layer (ablator)of a spherical assembled deuterium-tritium (DT) fuel capsule inside the hohlraum, resulting in the fusion fuel implosion compression and a center hotspot for ignition. In the past years, the ID hotspot ignition experiments, where the capsule is put inside a cylindrical hohlraum, have performed at National Ignition Facility (NIF) [3-6] and shown that the maximal radiation ablation pressure $P_a \sim 100$ Mbar at the radiation ablation front (RAF) for radiation temperature $T_r \sim 300$ eV [3, 6] is insufficient to drive the implosion compression, resulting in the convergence ratio (the initial outer radius of the capsule to the hotspot radius at the ignition time) $C_r > 30$ at stagnation (the hotspot stops motion inward) time and severe hydrodynamic instabilities without hotspot ignition so far[7].

As for the direct drive (DD) approach [2, 8], laser energy is directly absorbed near the critical surface and converted to an electronic-thermal (heat-conduction) wave (ETW) that propagates toward the electron ablation front at ablator surface, where the electron ablation pressure drives implosion compression of the capsule and the hotspot ignition. Although DD approach has high efficiency of laser to the capsule [9], hydrodynamic instabilities caused by the nonuniformity from laser beam overlapping near the capsule surface [10], and the lasers with high intensity interacting with the corona plasma [11,12] remain challenges.

In order to improve the ID and DD approaches, the recent proposed hybrid drive (HD) approach [13-15], which utilizes the ID and DD advantages and basically overcomes their shortcomings, offers a thermally-smoothed HD pressure $P_{HD}$, far higher than the radiation ablation pressure $P_a$ in traditional ID approach. A strong spherical-symmetric HD shock driven by $P_{HD}$ suppress hydrodynamic instabilities caused by the asymmetric ID shock that pre-compressed the capsule. Then, the HD shock further compresses the fuel and offers the hotspot ignition before stagnation (nonstagnation) time.

In this letter, a robust HD ignition target that involves a *spherical* hohlraun (SH)[16,17] and an assembled fusion fuel capsule inside SH is designed to expound how the maximal thermally-smoothed HD pressure as high as $P_{HD} \sim 650$ Mbar emerges by a novel effect and hydrodynamic instabilities in HD are suppressed by the strong spherically-symmetric HD shock driven by $P_{HD}$. Then, we will discuss the well-fitted scales of the maximal HD density and the maximal HD pressure, which are found in numerical simulations for different size targets and laser energies and confirmed experimentally on SG-III laser facility.

In the robust ignition target, SH is of radius $R_{SH} = 5$ mm and six laser-entrance holes (LEHs), as schematically plotted in Figs 1a, and the capsule is made up of the CH ablator (the outer radius $R_{ca} = 916$ μm) and the main fuel (0.23mg DT) layer with a central low-density fuel gas, as shown in Fig. 1b. Laser energy involving ID laser energy $E_{ID} = 0.65$ MJ with two pulses (two ID shocks) of duration of 7.5 ns and DD laser energy $E_{DD} = 0.768$ MJ with the power 320 TW of 2.4-ns flattop pulse, is delivered on target to drive HD implosion dynamics and hotspot ignition.

In whole implosion process, the ID lasers (schematically indicated by blue lines in Fig. 1a) act on the gold-shell wall of SH and its energy is converted to radiation temperature $T_r$ as seen in Fig.1c (black color). And the DD lasers beginning at t=5.1 ns, as shown in Fig. 1c (red color), propagate toward the capsule (schematically indicated by red lines in Fig. 1a) to be absorbed near the critical surface (indicated by red dished color in Fig. 1a) in the ID corona plasma produced by radiation ablating CH.

Using one-dimensional radiation hydrodynamics code-LARED [18], the main simulations results are plotted in Fig. 2. In the first implosion stage till t=5.1 ns, only ID radiation temperature $T_r$ ablates the CH ablator surface of the capsule. A high foot with $T_r$=120 eV at t~1 ns, see Fig. 1c, will offer a high adiabat for the compressed fuel. In the end of the first stage (t=5.1 ns), the radiation ablation pressure in the ablating CH surface of radius R~780 μm reached $P_a$~ 40 Mbar (red color in Fig. 2b). The capsule is pre-compressed to densities of ~ 2 g cm$^{-3}$ for DT and ~ 5 g cm$^{-3}$ for the unablated CH ablator (black color in Fig. 2b) by the first ID shock driven by the radiation ablation pressure. During the first stage, a long-scale-length ID corona plasma, which offers a background for HD implosion, as shown in Fig. 2a, is formed by the ablated CH ablator. In the second implosion stage (HD), from t = 5.1 ns (the launch time of the DD lasers) to 7.5 ns, the ID radiation continues ablating the CH ablator till t = 7.5 ns, together with the DD lasers to perform HD implosion dynamics. At t= 6 ns, $T_r$ rose to the peak of 190 eV, which further enhances the ID pre-compression of the DT fuel and offers more ablated mass to the ID corona plasma. And the second ID shock caught up with the first ID shock to form a new ID shock near the inner interface of the pre-compressed DT fuel.

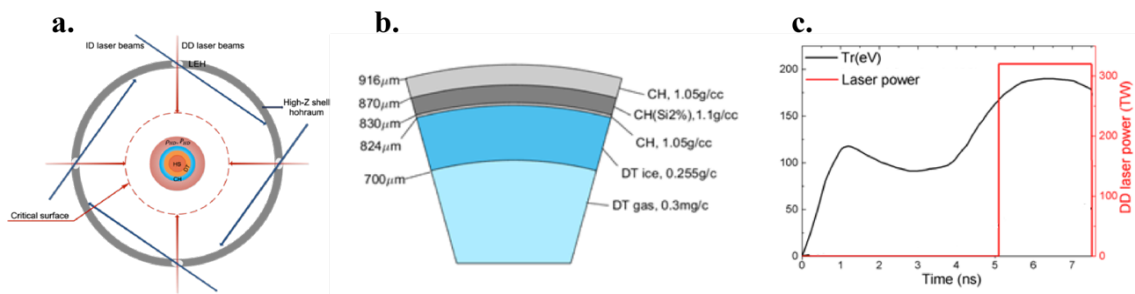

Fig. 1. Target configuration and laser energy for implosion. (a) The plane figure of the spherical hohlraum with six LRHs for ID (blue lines) is schematically plotted, where the DD lasers (dark-red lines), the critical surface (red broken line）, and the plateau region of $\rho_{HD}$ and $P_{HD}$ (dark-red region) are indicated. (b) The assembled DT fuel layered capsule. (c) Radiation temperature (black curve) converted by ID laser energy (tow pulses) and the DD laser power with 2.4-ns flattop pulse (red color).

At t = 5.1 ns, the DD lasers enter SH toward the ablating capsule, and then are absorbed and converted to the nonlinear supersonic ETW nearby the critical surface with radius $R_c \sim$ 1050 μm, which is ~250-300 μm away from RAF (see Fig. 2a). This supersonic ETW with an averaged velocity of ~ 625 km/s propagates within this distance toward RAF and at time t ~ 5.5 ns slows down to a plasma compression wave followed a precursor shock at R ~ 780 μm. The precursor shock compresses the DT fuel again and then catches up with the new ID shock at the inner interface of the fuel to generate a merged shock (MS) that compresses and heats the innermost part of the DT fuel to form a hotspot for ignition. The plasma compression wave, like the "bulldozer", steadily supported by the flattop DD laser pulse (red color in Fig. 1c), heaps low ID corona density into higher plasma density $\rho_{HD}$ to form a HD plasma density plateau (the grey color regions in Fig. 2b, also the dark-red region in Fig. 1a) between compression wave front (CWF) and RAF, where that's how the HD pressure appears. At t = 6.0 ns, the averaged HD density rosed to ~ 2.2 g/cc and the HD pressure was boosted to ~240 Mbar (red color in Fig.2b) in the HD density plateau ahead of RAF. Late, at t = 7.5 ns, RAF has moved inwardly to R ~ 300 μm, and the averaged HD density and HD pressure are ~3.1 g cm$^{-3}$ and ~486 Mbar respectively, see Fig.2b. At t~7.75 ns, the maximal HD density reached $\rho_{HD} \approx$ 4.25 g cm$^{-3}$ within the HD density plateau with width ~ 50 μm, where radiation temperature rose to $T_r \approx$ 230 eV larger than that 190 eV from ID lasers, which may come from additional electron bremsstrahlung and dopant Si radiation. Therefore, the maximal HD pressure is $P_{HD} \approx \Gamma \rho_{HD} T_r \approx$ 650 Mbar, where $\Gamma$ depending on the ablator materials is the constant for the ideal plasma. The maximal HD pressure drives a strong enough HD shock, which further compresses the DT fuel and at t ~ 7.68 ns rapidly arrives at the hotspot interface (radius R ~190 μm) with early convergence ratio of ~5, to collide with and stop MS that just first arrived there after reflected at the hotspot center. After stopping MS, the HD shock directly enters the hotspot that is further compressed and heated. Meanwhile, the DT fuel is continuously accelerated, and at t ~ 7.85 ns reaches its maximal value $V_{im} \approx$ 415 km/s with maximal fuel kinetic energy 20 kJ, which offers pdV work to the hotspot for ignition. The hotspot ignition occurs in nonstagnation time (t~ 8.06 ns) soon after the HD shock rebounded at the hotspot center reaches hotspot interface (HI). At that time, the velocity at HI is still centered in u ~ −180 km/s and the mechanical work $4\pi \int r_h p u dt$ till the stagnation time offers a margin about 25%. At ignition time, the power balance gives ignition conditions [14]: ion temperature $T_i \sim$ 5 keV (mass averaged) and the area density $(\rho R_h) \approx$ 0.2 g cm$^{-2}$ with the converged hotspot radius $r_h \sim$ 40 μm. Meanwhile, the hotspot pressure reached $P_{ig} \sim$ 200 Gbar with a convergence ratio $C_r \sim$ 23. It will greatly reduce the high-dimension risk and will be discussed later. Until the stagnation time at about 35ps after ignition, the ignited hotspot has release fusion energy ~ 200 kJ, resulting in a thermonuclear burning wave toward the high-compressed DT fuel. Finally, we achieved the fusion energy gain about 10. In whole implosion process, the capsule absorbs energy ~ 660 kJ with the efficiency of ~ 50% and the fuel adiabat α ~ 3.2, promising the HD capsule would be potentially robust to implosion instabilities.

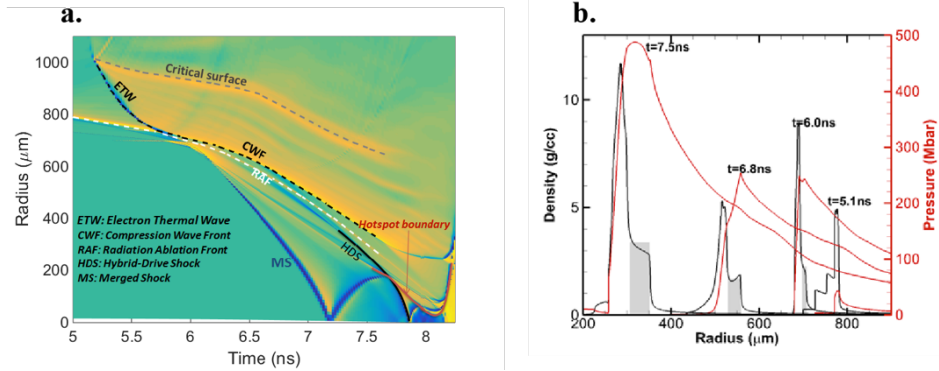

Fig 2. (Color online). (a) Radii of the critical surface, supersonic electron thermal wave, compression wave front, radiation ablation front, and ID and DD shocks in HD implosion vs time t. (b) Temporal evolution of the HD pressure (red) and density (black) of the imploding capsule. The density plateau (grey) is formed by the "bulldozer" effect.

In what follows, we will discuss high-dimensional impacts on HD implosion dynamics. In the traditional ID approach, high-dimensional impacts, which are mainly from thermal x-ray irradiating, supported capsule tent, and fill tube[19-21] as appeared in NIF experiments, give rise to asymmetries (nonuniformities) of the radiation ablation pressure $P_a$ and the ID shock driven by $P_a$, and hence cause severe hydrodynamic instabilities, resulting in obstacles of the hotspot ignition [3-7].

In HD, asymmetric implosions are mainly caused by nonuniformities of the radiation ablation pressure $P_a$ and the HD pressure $P_{HD}$. The asymmetry $\delta P_a/P_a$ mainly comes from two factors. One is the asymmetric flux $F_x$ of thermal x-rays irradiating on the ablator surface, which mainly is the Legendre $P_4$ for SH with six LEHs [14,15]. For the present target of $R_{SH}/R_{CA}$ = 5.46, $\delta P_a/P_a \sim |(\delta F_x)/F_x| \approx 0.45\%$ is easily obtained by view factor calculation. Another is the ablated tent that collapses on the ablator surface and emerges in a complex space distribution surrounding initial two touched locations.

We first of all discuss hydrodynamic instabilities caused by asymmetry $\delta P_a/P_a$. The growth factor (GF) = $\varepsilon(t)/\varepsilon(0)$ is used to investigate hydrodynamic instabilities at interfaces in the HD imploding capsule, where $\varepsilon(t)$ is the perturbed amplitude at time t and $\varepsilon(0)$ is the initial perturbation seeded on the capsule surface. In 2D simulations, the initial perturbation of the nonuniformity of x-ray irradiation at the ablator surface is multimode, while the tent collapsed material distribution is fixed by an approximation gaussian one (mainly in the range of modes n = 20 – 30) with the peak of 0.1μm. The simulations of GFs with modes n = 2 – 120 are plotted in Figs. 3 (a)and (b). At RAF, the maximal GF∼ +170 for n =

38 and minimal GF ∼−75 for n=80 are shown in Fig. 3(a). At the hotspot interface (or the internal interface of the high-compressed fuel), Gfs that mainly are feedthrough from RAF are particularly important for hotspot ignition. Our result shows that the peak is $(GF)_{hs}$ = −13 in mode n = 16, as plotted in in Fig. 3(c). All GFs in the present target (adiabat α = 3.2 and $C_r$ = 23) are far smaller than that in NIF experiments (α = 2.8 and $C_r > 30$)[6]. What the yield from 2D simulations is very close to the one-dimensional results shows that hydrodynamic instabilities in HD are neglectable.

We now discuss the nonuniformity of $P_{HD}$. It comes from the overlapping of the DD lasers that are absorbed near the critical surface, where the converted supersonic ETW has the pressure nonuniformity $(\delta P/P)_{cs} \approx (2/3)\delta I_L/I_L$. The laser intensity nonuniformity $\delta I_L/I_L \approx 5\%$ is calculated by the 3D 'ray-tracing' code for the present target. Thanks to the long enough distance ∼ 250-300 μm between the critical surface and RAF, it makes the nonuniformity of $P_{HD}$ to be well smoothed to a low level when the supersonic ETW slows down to the compression wave. Simulations by 2D radiation hydrodynamics code show that the relative ratio σ(R) = (δP/P)/(δP/P)$_{cs}$ quickly descends to ∼ 0.5% at the distance of 180μm away from the critical surface, and then, slowly decreases till RAF, as plotted in Fig. 3c. Therefore, the nonuniformity of $P_{HD}$ at RAF is < ∼0.3% and the HD shock driven by $P_{HD}$ is very close to spherical symmetry.

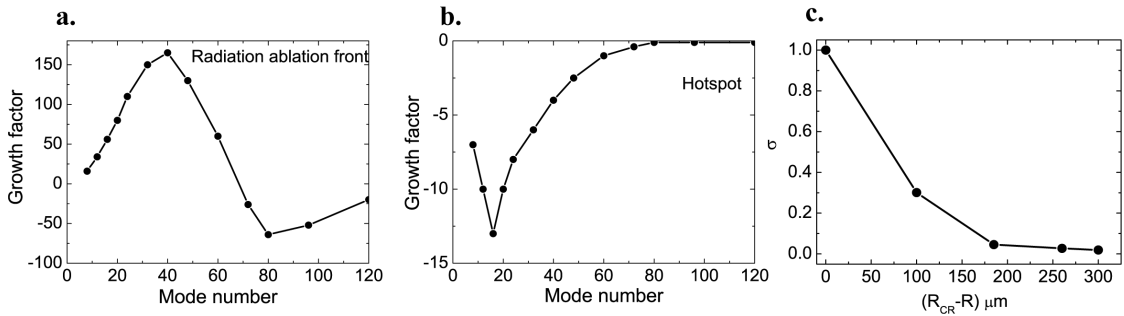

Fig. 3. (Color online). 2D impacts on implosion dynamics. (a) Growth Factor (GF) vs mode number and density profile at RAF with initial roughness of 0.1μm in the peak implosion velocity time at t ∼ 7.76ns. (b) GF at HI in the ignition time. (c) Pressure smoothing effect σ (black dots from 2D simulations) vs the distance $R_{cr} − R$.

The strong HD shock with good spherical symmetry and uniformity driven by the smoothed $P_{HD}$ quickly moves to HI, where the asymmetrically merged ID shock is stopped in early convergence ratio of ∼ 5 at t=7.68 ns, as mentioned in one-dimensional simulations. Then, the HD shock enters the hotspot that is further compressed and heated in one-dimension like. The hotspot ignition happens at the convergence ratio $C_r$ ∼23 and the fuel adiabatα = 3.2 without severe development of the 2D impacts when the HD shock for the first time reflects at HI after rebounded at the center.

Finally, we discuss the fill tube impact on the hotspot ignition by 2D simulation. The results show that a jet with low density material from the ablator along with the ID shock appears nearby the tube hole of initial diameter 10nm as the ablator surface is ablated by ID temperature $T_r$, and quickly enters the tube and reaches the DT gas region in the capsule till the hole is closed by the HD shock at t ~ 5.7 ns. Finally, the hotspot is mixed by additional materials, about 14ng at ignition time, from the CH ablator and the tube wall. As a result, the simulation shows that the mix effect on the yield is negligible.

The above discussions show that the 2D impacts on the HD implosion dynamics and hotspot ignition indeed are neglectable and implosions are in one-dimension like.

Using the radiation hydrodynamics codes-LARED series [18], we have done a lot of numerical simulations for different targets and laser energies. All the results reveal that as long as the capsule surface ablated by ID radiation offers a long distance of $\sim 250 - 300$ μm [14-15] between the critical surface and RAF, and peak radiation temperature in the second stage is $T_r \geq 160$ eV, the novel effect, "bulldozer" like, caused by the interaction of DD lasers with the ID corona plasma can significantly heap the HD density and boost the HD pressure within the plasma density plateau. We found that the maximal HD density $\rho_{HD}$ and hence the maximal HD pressure $P_{HD}$ driven by the novel "bulldozer" effect are of the well-fitted scales of $\rho_{HD}$ (g cm$^{-3}$) $= AE_{DD}^{1/4}$ and $P_{HD}$(Mbar) $= \Gamma\rho_{HD}T_r = BE_{DD}^{1/4}T_r$, where DD lase energy $E_{DD}$ is in unit of kJ, radiation temperature $T_r$ is in 100eV, and A and B are the constant depending on the ablator materials and A$\approx 0.85$ and B $\approx 56$ for the CH ablator. The scale relationship between maximal HD density $\rho_{HD}$ and DD lase energy $E_{DD}$ is plotted in Fig. 4a, where the red dot is for the present target discussed above. Simulations show that either $P_{HD} \geq 450$ Mbar and $T_r \geq 160$ eV or $E_{DD} \geq 650$ kJ and $E_{ID} \geq 500$ kJ for the SH radius $R_{SH} = 5$ mm is requested for the hotspot ignition at the low convergence ratio $C_r \sim 22 - 23$. It is such $P_{HD}$, over 5 times the peak radiation ablation pressure $P_a$ in NIF experiments [3-5], that suppresses hydrodynamic instabilities and makes the ignited hotspot releasing enough fusion energy to offer a burning wave propagating in the high compressed fuel, resulting in the fusion energy gain $\geq 10$. The scales are still available even if $E_{DD}$ is quite low but $T_r \geq 160$ eV. We have confirmed the scales experimentally on SG-III laser facility and achieved $\rho_{HD} \approx 1.2$ g cm$^{-3}$ and $P_{HD} \approx 155$ Mbar [15] about 3.5 times radiation ablation pressure $P_a$ using $E_{DD}$ =3.6 kJ (~$1.8 \times 10^{15}$ w $\cdot$ cm$^{-2}$) and $E_{ID} = 43$ kJ that is converted to $T_r = 200$ eV in a CH plane inside the half-cylindrical hohlraum. The experimental result is quite consistent with scales, as seen in Fig. 4a (red star). It shows that as long as $T_r \geq 160$ eV, the the "bulldozer" effect to boost the HD pressure is quite significant even if the DD laser energy is reduced to $E_{DD}$~kJ level. In the above experiment, we also measured the laser backscattering fraction about $(4.5 \pm 1)\%$ involving stimulated Raman scattering (SRS) and stimulated Brillouin scattering (SBS) (Fig. 4b) and the hot electron energy fraction about 2% (Fig. 4c) in the range of the DD laser intensities of $(1-2) \times 10^{15}$ w cm$^{-2}$ [15]. Both the fractions in HD are about a third of that in the traditional DD laser plasma interaction without ID corona plasma background, as seen in Figs.4b and 4c, and the physical reason will be explained elsewhere.

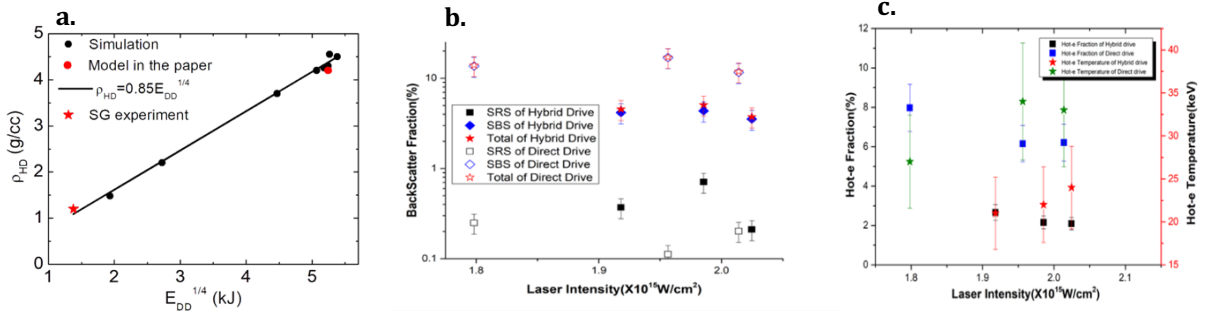

Fig. 4. (Color online). (a) The well fitted scales of hybrid-drive density $\rho_{HD} \approx A\, E_{DD}^{1/4}$ and hence $P_{HD} \approx B E_{DD}^{1/4} T_r$, the hotspot ignition is requested in $P_{HD} \geq \sim 450$ Mbar or $E_{DD}^{1/4} \geq 5$. The red dot is for the present target and the red star is for experimental result on SG-III. (b) The laser backscattering fraction (SRS+SBS) for HD and for the traditional DD without ID corona plasma background vs laser intensities. (c) The hot-electron energy fraction for HD and for the traditional DD without ID corona plasma background vs laser intensities

In conclusion, we report the HD implosion dynamics and the nonstagnation hotspot ignition. ID laser energy is converted to thermal X-ray radiation, in which the radiation ablation offers the long enough distance between the critical surface and RAF and the radiation ablation pressure $P_a$ results in the pre-compression of the assembled fuel capsule. While DD laser energy is absorbed near the critical surface. And the converted supersonic ETW propagating toward RAF in the long enough radiation-ablation distance slows down and then is converted into the "bulldozer effect, which results in the maximal smoothed HD pressure and the HD shock to suppress hydrodynamic instabilities caused by ID shocks and to perform nonstagnation hotspot ignition. As an example, the simulations of the robust ignition target show that the maximal HD pressure is as high as $P_{HD} \sim 650$ Mbar, the maximal implosion velocity reaches 415 km/s, and severe hydrodynamic instabilities are suppressed, resulting in nonstagnation hotspot ignition at the convergence ratio $C_r = 23$, and finally achieving the fusion energy gain~10 in total laser energy of 1.42 MJ. Two-dimensional simulations have confirmed that hydrodynamic instabilities are suppressed and the capsule implosion is in one-dimension like. The well-fitted scale between the maximal HD pressure (density) and the quarter power DD laser energy is found as long as $T_r > 160$ eV.

Experiments have verified that the "bulldozer" effect results in the boosted HD pressure about 3.5 times the radiation ablation pressure for CH ablator on SG-III using $T_r \approx 200$ eV($E_{ID} = 43$ kJ) and $E_{DD} = 3.6$ kJ only, also shown that both the laser backscattering fraction and the fraction of hot electron energy for the DD laser intensity of $\sim 1.8 \times 10^{15}$ w cm$^{-2}$ are about a third of that in traditional DD laser plasma interaction.

## Acknowledgments

Part of the work was supported by the National Natural Science Foundation of China under Grant Nos.11475032 and 11605178, Science Challenge Project (TZJH1616- 07), and the Science and Technology Developing Foundation of CAEP (Grant No. 2015B0202033).

## References

[1] J. D. Lindl, Phys. Plasmas 2, 3933(1995).

[2] L. Nuckolls et al., Nature 239, 139(1972).

[3] O. A. Hurricane et al., Nature 606, 343(2014).

[4] O. A. Hurricane et al., Phys. Plasmas 21，056314(2014).

[5] H.-S. Park et al., Phys. Rev. Lett. 112，055001(2014).

[6] T. R. Dittrich et al., Phys. Rev. Lett. 112，055002(2014).

[7] D. S. Clark et al., Phys. Plasmas 26, 050601(2019).

[8] R. Betti et al., Phys. Rev. Lett. 98, 155001(2007).

[9] S. Atzeni and J. Meyer-ter-Vehn, *The Physics of Inertial Fusion* (Oxford University Press, oxford, 2004).

[10] R. Betti and Q. Hurricane, Nature Phys. 12, 435(2016).

[11] W. Theobald et al., Phys. Plasmas 19, 102706(2012).

[12] D. Batani et al., Nucl. Fusion 54, 054009(2014).

[13] X. T. He et al., The updated advancements of inertial fusion program in China, in 8th Int. Conf. on Inertial Fusion Science and Applications, Sept. 8-13, 2013, Nara, Japan. Published in Journal of Physics: Conference series 688, 012029 (2016).

[14] X. T. He et al., Phys. Plasmas 23, 082706(2016).

[15] X. T. He, High Energy Density Physics, 100804 (2020).

[16] K. Lan et al., Phys. Plasmas 21, 010704(2014).

[17] K. Lan et al., Phys. Plasmas 21, 052704(2014).

[18] Z. F. Fan et al., Phys. Plasmas 21, 100705(2014).

[19] D. S. Clark et al., Phys. Plasmas 26, 050601 (2019).

[20] B. M. Haines et al, Phys. Plasmas 24, 072709(2019)

[21] B. M. Haines et al, Phys. Plasmas 26, 012707(2019).

[22] L. F. Wang et al., Phys. Plasmas 2, 122710(2014).